\documentclass[fleqn,usenatbib]{mnras}

\usepackage{graphics,epsf}
\usepackage{amsmath}                
\usepackage{amsfonts}               
\usepackage{amssymb}                
\usepackage{epsfig}                 
\usepackage{graphicx}

\usepackage{xcolor}
\definecolor{redak}{rgb}{0.9,0.15,0.05}

\def \s{~\rm{s}}
\def \km{~\rm{km}}

\def \erg{~\rm{erg}}

\def \days{~\rm{days}}

\def \rmModot{~\rm{M_{\sun}}}
\def \rmRodot{~\rm{R_{\sun}}}

\def \rmMJ{~\rm{M_J}}
\def \rmRJ{~\rm{R_J}}

\title[An ILOT model for the YSO ASASSN-15qi]{An intermediate-luminosity-optical-transient (ILOT) model for the young stellar object ASASSN-15qi}

\author[A. Kashi and N. Soker]{
Amit Kashi$^{1}$\thanks{E-mail: \href{mailto:kashi@ariel.ac.il}{kashi@ariel.ac.il}}
and
Noam Soker$^{2}$\thanks{E-mail: \href{mailto:soker@physics.technion.ac.il}{soker@physics.technion.ac.il}}
\\
$^{1}$Physics Department, Ariel University, Ariel, POB 3, 40700, Israel \\
$^{2}$Deparment of Physics, Technion, Haifa 3200003, Israel\\
\\
}

\date{Accepted XXX. Received YYY; in original form ZZZ}

\pubyear{2016}

\begin{document}
\label{firstpage}
\pagerange{\pageref{firstpage}--\pageref{lastpage}}
\maketitle

\begin{abstract}
We construct a scenario where the outburst of the young-stellar-object ASASSN-15qi is an intermediate luminosity optical transient (ILOT).
In this scenario a sub-Jupiter young planet was tidally destructed on to a young main-sequence star.
The system is young, therefore the radius of the planet is larger than its final value, and consequently its density is smaller.
The lower density allows the tidal destruction of the young Saturn-like planet on to the main-sequence star of mass $\approx 2.4 \rmModot$,
resulting in a formation of a disc and a gravitationally-powered ILOT.
Unlike the case of the more energetic ILOT V838~Mon, the mass of the destructed planet is too low to inflate a giant envelope, and hence the merger remnant stays hot.  
If our suggested model holds, this ILOT possesses two interesting properties:
(1) its luminosity and total energy are below those of novae, and (2) it is not as red as other ILOTs.
The unusual outburst of ASASSN-15qi, if indeed is an ILOT, further increases the diversity of the already heterogeneous group of ILOTs.
We mark the region on the energy-time diagram occupied by such young ILOTs.
\end{abstract}


\section{INTRODUCTION}
\label{sec:intro}

\subsection{General}
\label{subsec:general}

As in recent years the quality and quantity of sky surveys increase, more attention is given to rare explosions and outbursts in the energy gap between novae and supernovae
(e.g. \citealt{Mouldetal1990, Rauetal2007, Ofeketal2008, Ofeketal2016, Prietoetal2009, Botticella2009, Smithetal2009, Berger2009a, Berger2009b, KulkarniKasliwal2009, Mason2010, Pastorello2010, Kasliwaletal2011, Tylendaetal2013, Kasliwaletal2011, Kurtenkovetal2015, Tartagliaetal2016, Villaretal2016, Blagorodnovaetal2017}).
Those outbursts, known as intermediate luminosity optical transients (ILOTs) form an extended family that has a number of subgroups (see \citealt{KashiSoker2016} for a detailed nomenclature: Intermediate-Luminous Red Transients, LBV giant eruptions and SN Impostors, and Luminous Red Novae or Red Transients or Merger-bursts).  
 
Researchers have been modeling ILOTs, or subgroups of ILOTs, either as single-star phenomena (e.g., \citealt{Thompsonetal2009, Kochanek2011} for eruptive red giants and \citealt{Ofeketal2013} for a SN impostors), or as interacting binary systems  (\citealt{Kashietal2010, KashiSoker2010b, SokerKashi2011, SokerKashi2012, SokerKashi2013, McleySoker2014, Nandezetal2014, Goranskijetal2016, Pejchaetal2016b, Soker2016}), including a common envelope evolution \citep{RetterMarom2003, Retteretal2006, Tylendaetal2011, Ivanovaetal2013, IvanovaNandez2016, Tylendaetal2013, Nandezetal2014, Kaminskietal2015b, Soker2015, MacLeodetal2016, Blagorodnovaetal2017}.

\begin{figure*}
\centering
\includegraphics[trim= 0.8cm 0.1cm 1.5cm 0.5cm,clip=true,width=1.0\textwidth]{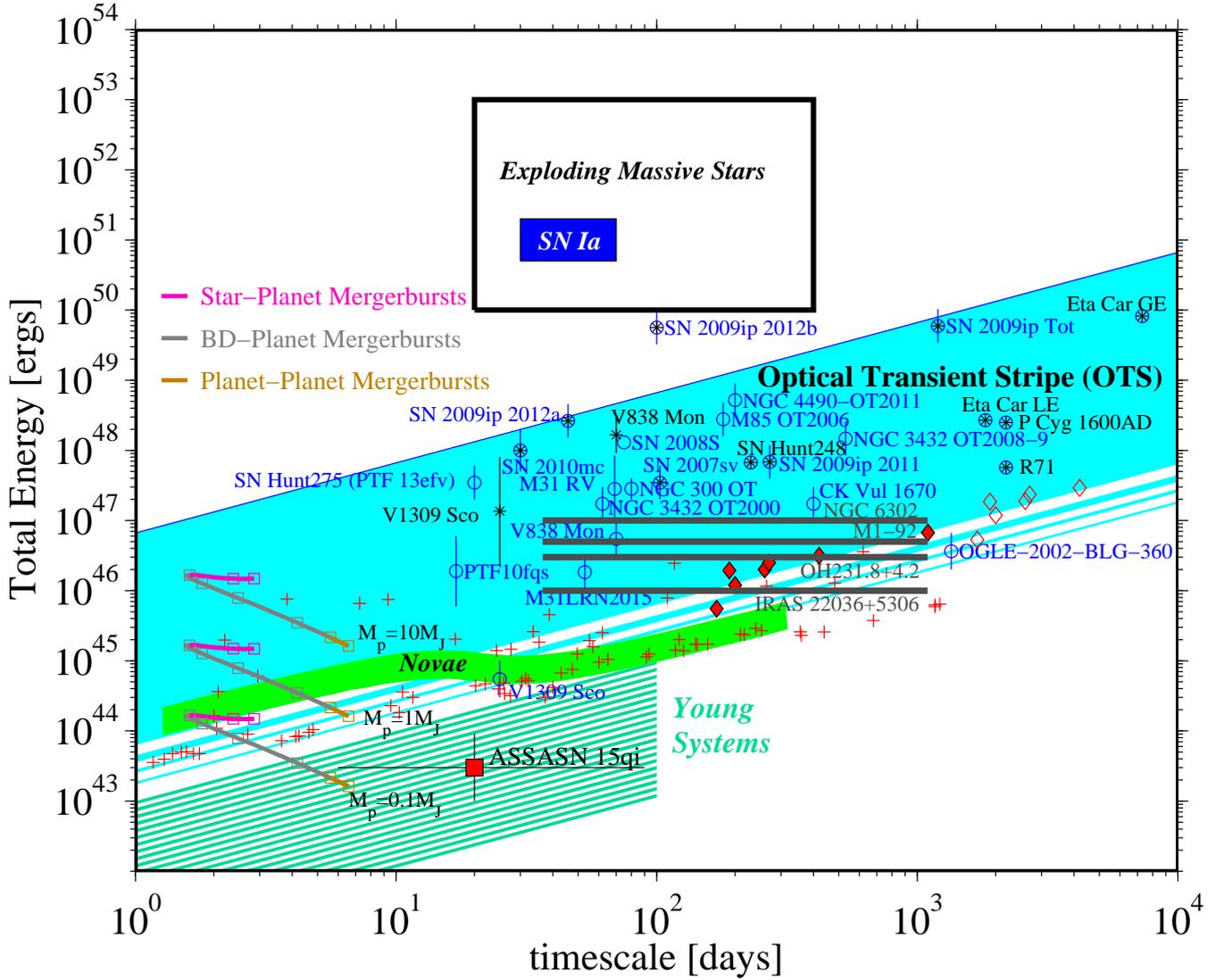}
\caption{
Observed transient events on the energy time diagram.
Blue empty circles represent the total (radiated plus kinetic) energy of the observed transients as a function of the duration of their eruptions, i.e., usually the time for the visible luminosity to decrease by 3 magnitudes. The Optical Transient Stripe is populated by ILOT events that we suggest are powered by gravitational energy of complete merger events or vigorous mass transfer events (\citealt{KashiSoker2010b, KashiSoker2016, SokerKashi2016}). 
For ILOTs that  had sufficient data to create a model to  calculate their total available energy, we mark it by a black asterisk above, or overlapping with, the blue circle.
The total energy does not include the energy that goes to lifting the envelope and does not escape from the star.
Novae models are marked with a green line \citep{dellaValleLivio1995}, with red crosses \citep{Yaronetal2005}, or with diamonds \citep{Sharaetal2010}.
The four horizontal lines represent planetary nebulae (PNe) and pre-PNe that might have been formed by ILOT events \citep{SokerKashi2012}.
Merger models of a planet with a planet/BD/star \citep{Bearetal2011} are shown on the left hand side, together with models we added for smaller merging planet with mass $0.1 \rmMJ$.
The lower-left part (hatched in green) is our \it{new extension for younger objects}, including ASASSN-15qi (red square), where the planets are of lower density and can more easily undergo tidal destruction. 
}
\label{fig:etd}
\end{figure*}

There are two main diagrams to characterize ILOTs. One is the peak luminosity versus eruption duration (time scale; \citealt{Rauetal2009,Kasliwal2013}), and the second diagram is that of the total eruption energy versus eruption time. The latter is called the energy-time diagram\footnote{An updated version of the energy time diagram is available at \url{http://phsites.technion.ac.il/soker/ilot-club/}}, and we present it in Fig. \ref{fig:etd}. 

In the present study we deal with a subgroup of ILOTs that are powered by gravitational energy which is released from a complete merger process of two stars (termed Luminous Red Novae, or Red Transients, or Merger-bursts), such as V838~Mon \citep{SokerTylenda2003, TylendaSoker2006} and V1309~Sco \citep{Tylendaetal2011, Nandezetal2014, Kaminskietal2015a}.
Differently from the objects above, in this study we discuss the destruction of a planet on a star, rather than a star on a star.  

\subsection{Star-Planet Mergers}
\label{subsec:merger}

\cite{Bearetal2011} propose that a V838~Mon like merger-burst can happen on smaller scales and low energies -- between a planet and a low mass main-sequence (MS) star, between a planet and a brown dwarf (BD) or between two planets.
In this process the planet is tidally shredded into a disc, and the accretion of the gas in the disc onto the star, onto a brown dwarf, or onto another planet, leads to an outburst.
According to this model, the destruction of the planet occurs before it touches the more massive object, because the density of the planet is lower than the density of the more massive object. 
For a typical mass of the destroyed planet of $\approx 1 \rmMJ$, these outbursts populate the lower left part of the optical transient stripe, with timescales of a few days and total energies of $10^{44.3}$--$10^{46.3} \erg$ \citep{Bearetal2011}.
We note that in the triple-planet scenario proposed by \cite{RetterMarom2003} and \cite{Retteretal2006} for the outburst of V838~Mon, the planets enter the envelope of the star intact, and hence their scenario is different than the scenario we propose here.    
  
The thorough study by \cite{Metzgeretal2012} further established the star-planet merger process as member of the ILOT heterogeneous group.
They study the interaction between a Sun-like star and planets of masses of $1 \textrm{--} 10 \rmMJ$, and find that the ratio of the mean densities of the planet and the star determines the outcome.
For low enough mean density of the planet, the interaction can lead to tidal-dissipation event where the planet transfers mass at about steady rate to the star.
For density ratio in the range $\approx 1$--$5$, \cite{Metzgeretal2012} find that the mass transfer is unsteady, resulting in a dynamical disruption of the planet into an accretion disc around the star.
\cite{Metzgeretal2012} further calculate  the luminosity of such a transient to be in the order of $\approx 10^{37} \erg$ and the time scale to be in the order of a few weeks.

\subsection{The transient  ASASSN-15qi}
\label{subsec:ASASSN}

\begin{figure*}
\centering
\includegraphics[trim= 0.0cm 0.1cm 0.5cm 0.0cm,clip=true,width=1.0\textwidth]{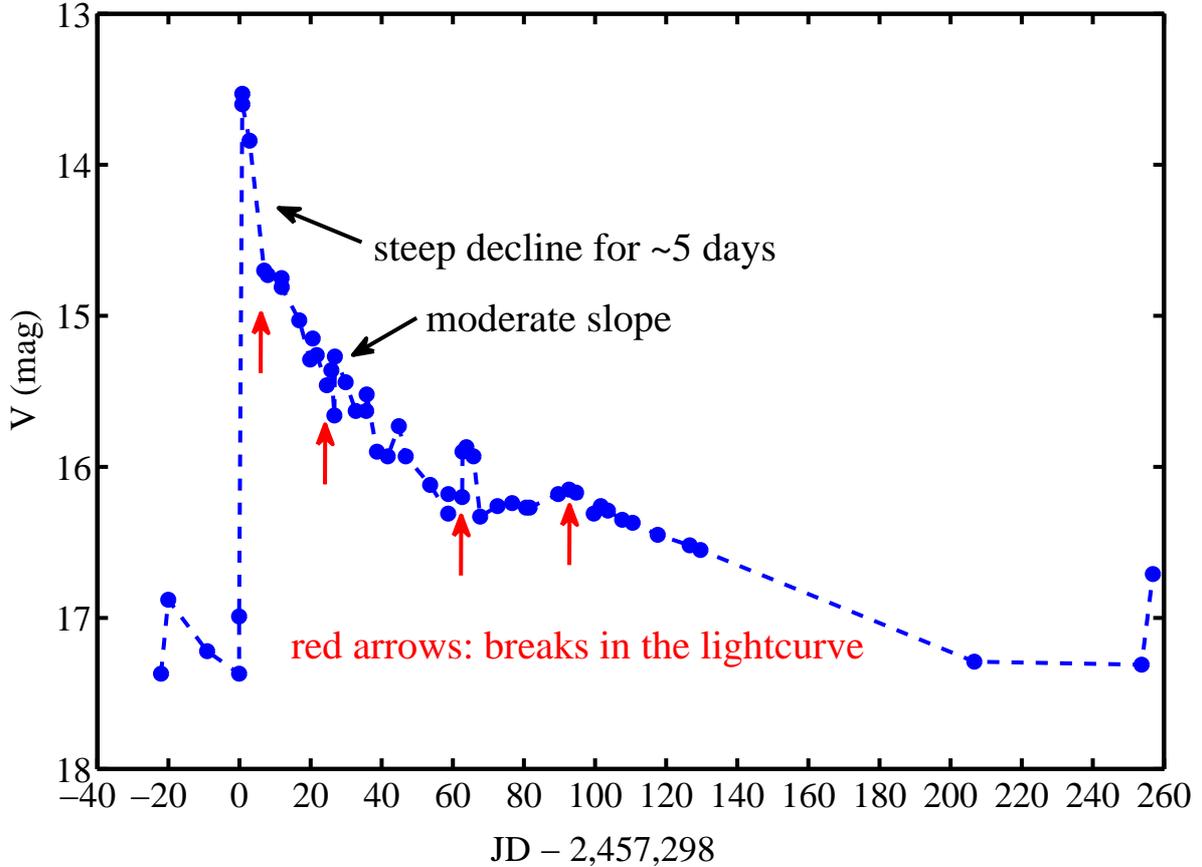}
\caption{
The light curve of ASASSN-15qi from the observations in Herczeg et al. (2016).
A change from a sharp decline to a moderate slope occurs about 5--6 days after the peak.
The red arrows indicate times at the light curve where a break in the slope of the light curve appears.
}
\label{fig:lightcurve}
\end{figure*}

The All-Sky Automated Survey for Supernovae (ASAS-SN) variability survey \citep{Shappeeetal2014}
discovered an outburst designated ASASSN-15qi (also referred to as 2MASS J22560882+5831040) on JD 2,457,298 (2 October 2015).
\cite{Herczegetal2016} report the observational properties of ASASSN-15qi in detail.
We present the light curve in V in Fig. \ref{fig:lightcurve}.
The following make the ASASSN-15qi outburst interesting.
(1) Its location among young objects, suggesting it is associated with a young stellar object (YSO).
(2) It showed a fast brightening of 3.5 mag in the optical bands in less than 23 hours.
(3) It blew a fast wind that faded as the outburst decayed over 4-5 months.

\cite{Herczegetal2016} calculate the outburst radiative energy to be $E_{\rm rad} \approx 7 \times 10^{42} \erg$ over a duration of 6 months. The kinetic energy might be much larger than the radiated energy, and it might accounts for most of the energy of the outburst. \cite{Herczegetal2016} mention that other outbursts from young stellar objects, such as V899~Mon and Z~CMa, share some similar spectroscopic features with ASASSN-15qi.
They speculated that ASASSN-15qi might be either a mass transfer event, connected to interactions between a star and a planet on an eccentric orbit, formation of an
excretion disc, or be some kind of a magnetic reconnection and outflow event.

The energy and timescale of ASASSN-15qi place this event on the energy-time diagram just below the region where \cite{Bearetal2011} predict the location of events where a merger of a planet with a low-mass main-sequence star take place. 
This close location motivates us to propose a planet-destruction model for the intermediate luminosity optical transient ASASSN-15qi.

In addition to the event in 2015, \cite{Herczegetal2016} mention an earlier outburst in 1976. We here concentrate on the event of 2015. If our scenario holds, then two planets have been destructed on the star, one in 1976 and one in 2015. Surely two such events make our proposed scenario much rarer. Yet, there are now many known planetary systems with a large number of planets and with rich variety of properties. Most pronounced is the detection of 7 earth-like planets around a very low mass star \citep{Gillonetal2017}. Our proposed scenario requires a planetary system with several planets, as one or more planets should perturb the orbit of the planet to be destroyed. We note that a scenario for an ILOT as a result of triple-planet collisions have been proposed by  \cite{Retteretal2006} for the outburst of V838~Mon. 

\section{A PLANET-DESTRUCTION EVENT}
\label{sec:model}

We propose that the young stellar object outburst ASASSN-15qi is an ILOT event, and construct a planet-destruction scenario where a sub-Jupiter young planet was tidally destroyed onto a young main sequence star.
If our proposed scenario holds, this ILOT is unusual in two aspects.
First it is not `intermediate' between the typical luminosities of novae and luminosities of supernovae, because its luminosity is like those of novae.
Second, it does not have a red photosphere as is the case for other ILOTs that are not LBV major outbursts, though it will appear red as a result of extinction. Below we explain that this non-red photosphere is the result of not enough mass being available to inflate a giant envelope. 

The density ratio of the planet to that of the star determines the outcome of the merger \citep{Metzgeretal2012}. We here study a case where the planet is destroyed outside the star, hence the planet density should be lower than the density of the star. This in turn requires the planet to be young and be larger than its final equilibrium size. 

A Saturn like planet ($M_p\simeq 0.3 \rmMJ$), for example, has a radius of $R_p \simeq 0.95 \rmRJ$ at an age of $300$ Myr (e.g., \citealt{Fortneyetal2007}),
compared to its final radius of $\simeq 0.84 \rmRJ$.
The radii of young planets depend on many parameters. The important ones are the planet composition, the size of the core, the accretion rate onto the core, the planet pressure profile, and the efficiency of cooling with the presence of stellar irradiation and tidal heating (e.g., \citealt{Guillot2005} and references therein).
In a recent review \cite{Baruteauetal2016} present some modern calculations of young planets, with a wide range of parameters that determine their radii.
At a very young age of $30$~Myr these radii are in the range  
\begin{equation} 
R_p \simeq (1.2 \textrm{--} 2.1) R_{pf},  
\label{eq:Rp}
\end{equation}
where $R_{pf}$ is the final radius of the planet. 
We shall use the most pessimistic value to our proposed scenario of $R_p = 1.2 \rmRJ$.

Solar type stars contract to their final zero-age main sequence radius in about their thermal timescale of $\approx 30$~Myr. More massive stars contract on a shorter time scale (e.g., \citealt{Bernasconi1996}). A star with a mass of $M_{\ast}=2.4 M_\odot$ contracts within a time scale of about only $\approx 5$~Myr. Moreover, the zero-age main sequence radius of stars is smaller than their radius at later times when they are still on the main sequence. For example, the solar zero-age main sequence radius was about 10 percent smaller than its value today (e.g., \citealt{Bernasconi1996}). The important point is that while the planet is larger than its final size in the relevant time for our proposed planet-destruction event, the central star had enough time to reach the main sequence. The stellar density is larger than the planet density at the considered time. 

The orbital separation at which planets are shredded by tidal forces is given by (e.g. \citealt{Nordhausetal2010})
\begin{equation}
\begin{split}
R_s &\approx R_p \left( \frac{2M_{\ast}}{M_p} \right)^{1/3} \\
    &  =  3 \left( \frac{R_p}{1.2 \rmRJ} \right)
         \left( \frac{M_{\ast}}{2.4 \rmModot} \right)^{1/3}
         \left( \frac{M_p}{ 0.3 \rmMJ} \right)^{-1/3} \rmRodot,
\end{split}
\label{eq:Rs}
\end{equation}
where $M_{\ast}$ is the mass of the star taken from \cite{Herczegetal2016}.
This tidal-destruction radius is larger than the radius of the star, and hence equation (\ref{eq:Rs}) implies that the young planet is tidally destroyed outside the star. This orbital separation at tidal destruction is larger than the tidal destruction separation at older ages.  
  
At the tidal destruction orbital separation given by equation (\ref{eq:Rs}) the Keplerian time is $t_{\rm K} \simeq 0.3$ days, which is well within the 23 hours constrain for the increase in luminosity of ASASSN-15qi \citep{Herczegetal2016}. If the planet has an eccentric orbit around the star with a pariastron distance smaller than $R_s$, then the tidal destruction process continues as the planet moves further in along its orbit. As a result of that the accretion process will be shorter even. An eccentric orbit might result from a perturbation by one or more larger planets. In general, if we are to explain also the 1976 outburst, we require that the systems contains several planets in unstable orbits.   

The mass accreted onto the star releases a gravitational energy that amounts to 
\begin{equation}
\begin{split}
E_{\rm acc} &\simeq 0.5 \frac{GM_\ast M_{\rm acc}}{R_{\ast}} \\
            &= 2.3 \times 10^{44} \left( \frac{M_{\ast}}{2.4 \rmModot} \right)
              \left( \frac{M_{\rm acc}}{10^{-4} \rmModot} \right)
              \left( \frac{R_{\rm ast}}{2\rmRodot} \right)^{-1}      \erg  .  
\end{split}
\label{eq:Ldrag}
\end{equation}
In scaling equation (\ref{eq:Ldrag}) we used the well studied ILOT V838 Mon. 
\cite{SokerTylenda2006} estimate the kinetic energy of the ejected mass in V838~Mon to be $E_{\rm kin} \approx 10^{44} \erg$, about an order of magnitude larger than the radiated energy. 
The ejected mass in ASASSN-15qi could have carried a similar amount of kinetic energy. For example, an ejected mass of $M_{\rm ej} \approx 0.1 M_{\rm acc} \approx 10^{-5} M_\odot$ that is ejected at the observed velocity of $v \approx 1000 \km \s^{-1}$ \citep{Herczegetal2016} has a kinetic energy of $E_{\rm kin} = 10^{44} \erg$. 

As suggested by \cite{Bearetal2011}, the accreted mass may form an accretion disc or an accretion belt around the star.
The disc that is formed by such a violent process is not expected to be flat. We will therefore scale with $H/R_a=0.3$, where $H$ is the thickness of the disc at a distance of $R_a$ from of the center of the star.
The accretion time $t_{\rm acc}$ of mass from the accretion disc should be longer than the viscous time scale for the gas in the disc to lose its angular momentum
\begin{equation}
\begin{split}
t_{\rm acc} & \ga t_{\rm{visc}} \simeq \frac{R_a^2}{\nu} \simeq 7
\left(\frac{\alpha}{0.1}\right)^{-1}
\left(\frac{H/R_a}{0.3}\right)^{-1} \\
   &   \times  \left(\frac{C_s/v_\phi}{0.3}\right)^{-1}
 \left( \frac{R_a}{3\rmRodot} \right)^{3/2}
\left( \frac{M_{\ast}}{2.4 \rmModot} \right)^{-1/2} \days,
\end{split}
\label{eq:tvisc1}
\end{equation}
where in the equation above $C_s$ is the sound speed, $\alpha$ is the
disc viscosity parameter, $v_\phi$ is the Keplerian velocity, and $\nu=\alpha ~C_s H$ is the viscosity of the disc.

The light curve of ASASSN-15qi shows that for about five days the decline in luminosity is very steep. Only after five days the light curve decline slope becomes moderate. This might be related to the viscosity time scale of the disc as we derived in equation (\ref{eq:tvisc1}). 

Most ILOTs have a red photosphere after the outburst because the envelope expands to large dimensions. In those cases the envelope of the merger product has a structure of an asymptotic-giant-branch (AGB) star. In V838~Mon, for example, the gas of the destroyed low mass main sequence star inflated a giant envelope with an envelope mass of $M_{\rm env} > 0.01 M_\odot$ \citep{TylendaSoker2006}. 

The ASASSN-15qi event was not red. In our proposed model of tidally destroyed planet, the bluer event is explained by the small mass of the destroyed planet. Only a very small mass was potentially available to inflate an envelope and make the star a giant.
Although the luminosity of ASASSN-15qi is similar to an AGB or an upper red-giant-branch (RGB) star, the envelope mass was too low to inflate an extended envelope.
When the envelope mass of AGB stars decreases to $M_{\rm env} \la 0.1 \rmModot$ they cease to expand. When their envelope mass further decreases to several$\times 10^{-4} \rmModot$ the envelope radius decreases to only several~$\rmRodot$ (e.g., \citealt{Soker1992}). 
As mass  is unavailable to  form an envelope, the merger product stayed small and hot, and not much dust was formed.

To summarize this point, this ILOT, although might be categorized as a luminous red nova by the properties of the merger process (see \citealt{KashiSoker2016}), did not become red since a low mass planet was the source of the accreted mass. 

An even more speculative scenario that might lead to similar ILOT properties can be constructed by using a more massive young planet, such as a young Jupiter-like planet. In this scenario the young Jupiter-like planet has an eccentric orbit that brings it very close to the star at periastron passages. Yet, this close distance is far enough that only the low-density outer layers of the planet are removed. Only a mass of $\approx 10^{-4} \rmModot$ is removed from the planet and accreted on to the star. The planet itself survives the encounter. This speculative suggestion should be studied in the future with a three-dimensional hydrodynamical code of binary interaction.     
In this scenario, the 1976 event was a previous periastron passage of the Jupiter-like planet, and this scenario predicts another outburst in 2054.

\section{POPULATING THE LOWER ENERGY-TIME DIAGRAM}
\label{sec:etd}

We now compare ASASSN-15qi with other ILOT events by placing it on the energy time diagram (Fig. \ref{fig:etd}). 

For the total energy of the 2015 event we take $E_{\rm tot}=10^{43} \textrm{--} 10^{44} \erg$, as we explained in section \ref{sec:model}. 
The energy of the ASASSN-15qi 2015 event is well below the energy of low-energy novae and below the optical transient stripe (marked in blue on Fig. \ref{fig:etd}). This goes also to the mean ratio of the luminosity to the Eddington luminosity of this event  
\begin{equation}
\Gamma_{\rm Edd} = \frac{(E_{\rm tot}/t) \kappa}{4 \pi G M_{\ast} c} \simeq 0.002 \textrm{--} 0.02, 
\label{eq:gamma_edd}
\end{equation}
where $\kappa$ is the opacity, and $c$ and $G$ have their usual meanings.
This ratio is much smaller that the typical values of $\Gamma_{\rm Edd,OTS} \simeq 0.1 \textrm{--} 10$ for other ILOTS on the optical transient stripe.

It is complicated to estimate the timescale of this ILOT.  
We usually take the ILOT timescale to be the time it takes the visible luminosity to decline by 3 magnitudes from its peak value. 
This criterion is based on the observations that in most ILOTs the radiated energy is emitted mainly in the V and R bands, and on the similar shapes of the time-scaled lightcurves of different ILOTs from the peak to $\approx 3 \textrm{--} 4$ magnitudes below the peak \citep{Kashietal2010}.

The shape of the lightcurve of ASASSN-15qi does not resemble the characteristic lightcurve of other ILOTs (as described in \citealt{Kashietal2010}). We cannot make the light curve similar even by rescaling the time, because the peak luminosity of ASASSN-15qi was only about 3 magnitudes above the stellar luminosity before and after the 2015 event. 
Furthermore, the lightcurve of ASASSN-15qi shows a few distinct breaks, at $\approx 6, 20, 60$ and $90 \days$ after the peak that are not seen in other ILOTs (see Fig. \ref{fig:lightcurve}.  

We notice the following interesting property of the lightcurve.
There is an exponential decay of 1 magnitude in 6 days, e.g., before the first break. 
There is a constant decay slope in the lightcurve after 90 days post-maximum.
If we extrapolate this part backward in time, we find that it intersects the original lightcurve at 20 days post-maximum, at the location of one of the breaks. 
This might indicate that the decline has two main phases, with a transition around 20 days post-maximum.
This complicated lightcurve hampers us from determining an accurate value for the decline time of this ILOT. We simply take this time scale to be in the range of $t=6\textrm{--}90 \days$, and mark ASASSN-15qi at a timescale of  $20 \days$ in the energy time diagram (Fig. \ref{fig:etd}). 
 
Following our proposed planet-star merger scenario for the 2015 event of ASASSN-15qi, we now extend the planet-merger region on the energy time diagram beyond the region presented originally by \cite{Bearetal2011}. This new region (hatched-green region in Fig. \ref{fig:etd}) requires young systems with ages of less than about $0.3$~Gyr. 

\section{SUMMARY AND DISCUSSION}
\label{sec:summary}

We examined the possibility that the unusual outburst of the YSO ASASSN-15qi is an ILOT event, similar in many respects to V838~Mon, but much fainter and of lower total energy.
As in the model of V838~Mon, the erupting system was young, but unlike the model for V838~Mon, we here suggested that the secondary object that was tidally destroyed onto the primary main-sequence star was a Saturn-like planet rather than another low mass main-sequence star. 

The young age in both systems is crucial. The reason is that along most of the main-sequence the density decreases with increasing stellar mass.
Therefore, low mass stars on the main-sequence will not suffer a tidal destruction by more massive primary main-sequence stars.
This holds for sub-stellar secondary objects as well. In young systems, the time scale on which the object shrinks to its final radius is longer for smaller mass.
Therefore, young low-mass objects have lower density than their final density, while the much more massive star can already be very close to its zero-age main sequence density.
The density of the low mass secondary object can be low enough to allow tidal destruction if it comes close enough to the primary star.

Very low-mass main-sequence stars and brown dwarfs in old systems have higher densities than more massive main sequence stars, and they can tidally destroy old planets, as well as young ones. This is indicated by the three lines in the lower left of Fig. \ref{fig:etd}.

The destruction of young planets on stars results in weak ILOTs, as we suggest here for ASASSN-15qi. Such events occupy the lower left part of the energy time diagram (hatched-green region in Fig. \ref{fig:etd}). The typical luminosity of these events is not necessarily above novae, and they do not have a red atmosphere. So they are not really intermediate between the luminosities of novae and supernovae, and they are not internally red.

The difference in the mass of the destroyed object between V838~Mon and ASASSN-15qi brings another significant difference in the properties of the merger product.
The luminosity of the outburst in both cases is in the range of giant stars.
In the case of V838~Mon, part of the mass of the destroyed secondary star inflated a huge envelope. In our proposed scenario for the 2015 outburst of ASASSN-15qi, the mass that is available to inflate an envelope is $<0.001 \rmModot$.
As is known from the evolution of post-AGB stars, this mass is too low to build a large envelope.
This is the reason that ASASSN-15qi had a relative small radius during the event, and hence it was relatively hot and not red. 

ILOTs constitute an heterogeneous group of objects that covers a large area in the energy-time diagram or the luminosity-time diagram.
We here suggested that this group, that is powered by gravitational energy in binary systems, is more heterogeneous and covers a larger area in these diagrams than the conventional view until now was.


\label{lastpage}
\end{document}